# Relationship between dielectric properties and structural long-range order in $(x)Pb(In_{1/2}Nb_{1/2})O_3:(1-x)Pb(Mg_{1/3}Nb_{2/3})O_3$ relaxor ceramics

C. W. Tai[*][†] and K. Z. Baba-Kishi

Department of Applied Physics and Materials Research Centre,

The Hong Kong Polytechnic University, Hung Hom, Kowloon, Hong Kong

The dielectric and structural order-disorder properties of as-sintered complex perovskite $(x)Pb(In_{1/2}Nb_{1/2})O_3:(1-x)Pb(Mg_{1/3}Nb_{2/3})O_3$ ceramics are highly influenced by the quantity of $Pb(In_{1/2}Nb_{1/2})O_3$ (PIN). High PIN quantity causes the relative permittivity maxima ($\varepsilon_{max}$) to decrease and the temperature ($T_{max}$) to increase. Also, strong frequency dispersion is dominant in the relative permittivity plotted against the temperature. In the ferroelectric hysteresis loop measurements, the maximum values of electric displacements ($D_{max}$) decrease with increasing PIN. The ceramics in the composition range $x = 0.1$ to $0.8$ behave as ferroelectric relaxors and exhibit very slim hysteresis loops for all these compositions. Transmission electron microscopy (TEM) studies show that the size of the 1:1 structural ordered domains is influenced by the PIN quantity. The relationship between the dielectric properties and the long-range 1:1 order in these relaxors appear to be in conflict with the commonly accepted order-disorder behavior in complex perovskite ferroelectrics, in which large structural domains correspond with the tendency to depart from the relaxor state. TEM observations show that individual 1:1 ordered domains in (x)PIN:(1-x)PMN ceramics are composed of numerous nano-sized ordered domains, separated by fine antiphase boundaries.

---

[*] Author to whom correspondence should be addressed. Email: cheukw.tai@gmail.com
[†] Present address: Department of Materials and Environmental Chemistry, Arrhenius Laboratory, Stockholm University, S-106 91, Stockholm, Sweden

## 1. Introduction

Lead-based complex perovskite-structured ferroelectrics with chemical formula $Pb(B'_xB''_{1-x})O_3$ are widely used as capacitors, actuators, sensors, electro-optical and MEMS applications [1,2,3]. In this class of materials, the two cations B' and B" often develop 1:1 long-range structural order on the {111} planes or may randomly occupy crystallographically identical B-sites positions. It is widely accepted that such order-disorder behavior influences the physical properties of these materials [4,5]. $Pb(Sc_{1/2}Ta_{1/2})O_3$ (PST) is a good example in which the influence of long-range order on properties is well documented [6,7]. A disordered or incompletely ordered complex perovskite normally exhibits board relative permittivity ($\varepsilon_r$), diffuse phase transition, strong frequency dispersion about dielectric maxima ($\varepsilon_{max}$) and a significant departure from the Curie-Weiss law. The order-disorder transformation can occur by thermal activation or cation substitution. Consequently, the physical properties can be changed by the influence of the degree of long-range order. Lead magnesium niobate, $Pb(Mg_{1/3}Nb_{2/3})O_3$ (PMN), is a well- known lead-based perovskite relaxor material. It exhibits strong frequency dispersion in $\varepsilon_r$ and slim ferroelectric hysteresis loop. PMN does not develop extensive long-range 1:1 structural order of the B-sites. The cation ordering cannot be enhanced by thermal treatment but can be changed by cation substitution [8]. However, the degree of structural order in $Pb(In_{1/2}Nb_{1/2})O_3$ (PIN), can be enhanced or reduced by thermal annealing and quenching [9]. PIN is antiferroelectric below the transition temperature in the ordered state whereas it is a relaxor in either incompletely ordered or disordered states.

$(x)Pb(In_{1/2}Nb_{1/2})O_3:(1-x)Pb(Mg_{1/3}Nb_{2/3})O_3$ ((x)PIN-(1-x)PMN), was first introduced by Lee and Kim (1998) [10]. Later, some of these solid solution ceramics were identified as ferroelectric relaxors [11]. In addition to the study of the dielectric permittivity, the room temperature ferroelectric hysteresis and electromechanical properties were also measured in a number of selected compositions [11,12]. The structural characteristics of (0.4)PIN:(0.6)PMN ceramic were studied by transmission electron microscopy (TEM) and high-resolution TEM [13,14]. In addition to the superstructure reflection 1/2 1/2 1/2, which is usually found in many complex perovskites, weak forbidden reflection 1/2 1/2 0 and transverse diffuse streaking were observed, indicating the presence of mixed cation order. However, an investigation of (x)PIN:(1-x)PMN ceramics in which the correlation between the physical properties and structural long-range order has not been reported. In this study, (x)PIN:(1-x)PMN ceramics with x = 0.1 to 0.9 are fabricated and their relative permittivity, ferroelectric hysteresis loops and ordered structural domains are investigated.

## 2. Experimental Procedure

(x)PIN-(1-x)PMN ceramics with x = 0.1 to 0.9 were prepared by the two-step solid state reaction [15]. The precursor materials were synthesized using lead (II) oxide (PbO, 99.9+%, Acros Organics Ltd.), magnesium (II) oxide (MgO, Reagent grade, Acros Organics Ltd.), indium (III) oxide ($In_2O_3$, 99.999%, Strem Chemical Inc.) and niobium (V) oxide ($Nb_2O_5$, 99.9%, Strem Chemical Inc.). The precursors were mixed together by ball-milling in ethanol with zirconia media for 6 hours. Columbite phase $MgNb_2O_6$ [15] and Wolframite phase $InNbO_4$ [16] were reacted at 1150 °C for 6 hours and 1200°C for 24 hours, respectively. The calcined powders were ball-milled again for 2 hours. The amounts of PbO, $InNbO_4$ and $MgNb_2O_6$ used follow the chemical formula. The 0.5 mole % of excess PbO was further added. The precursors were ball-milled in ethanol with zirconia media for 6 hours. The calcination temperatures varied from 850 °C to 880 °C, depending on the compositions of the ceramics. Polyvinyl alcohol (PVA) as a binder was added to the calcined powder. Pellets were obtained by uniaxial pressing. In order to reduce lead loss in the ceramics, a Pb-rich atmosphere was maintained during sintering. The ceramics were submerged in the powder of the same composition. The ceramics were heated up to the calcination temperature for one hour. The ceramics were sintered at 1200 °C for 2 hours [10].

The bulk densities of the sintered ceramics were measured by the Archimedes method. The densities of the ceramics with x ≤ 0.7 were obtained over 95 % of the theoretical values but the ceramics with x = 0.8 and 0.9 reached about 90 % and 85 % of the theoretical densities, respectively. X-ray diffractometry (Philips X'pert diffractometer) was used to determine the phases and crystal parameters of the sintered samples. The powder diffraction spectra were obtained from the crushed ceramics. Prior to the electrical measurements, the sintered ceramics were polished and coated with silver onto the surfaces as electrodes. Their $\varepsilon_r$ and dielectric loss (tan $\delta$) were measured as functions of temperature and frequency using a computer controlled analyzer (Alpha-A Analyzer, Novo-control Technologies). The samples were placed in a temperature-controlled cryostat system (Quatro Cryosystem, Novo-control Technologies), and were heated and cooled at a rate of 1 °C /min. The hysteresis loops of the ceramics were obtained at room temperature using a Sawyer-Tower circuit [17]. Specimens for the TEM study were mechanically polished to a thickness of 20 μm prior to ion-milling at 3.5kV (Gatan PIPS 691). The results were obtained with a JEOL-2010 transmission electron microscope operated at 200 kV and equipped with a double-tilt stage. The centered dark-field images were obtained using one of the superstructure reflections 1/2 1/2 1/2 along the [110] zone axis.

## 3. Observations

### 3.1 X-ray Diffractometry

Figure 1a is a series of x-ray diffraction patterns recorded from (x)PIN:(1-x)PMN ceramics with x = 0.1 to 0.9. The patterns show single-phase perovskite-structured ceramics with x ≤ 0.7. Evidence for the pyrochlore or other second phases was not detected in the patterns. Individual PMN or PIN peaks are not observed also, indicating the homogeneity of the solid solution formation. In the absence of peak splitting, all the reflections were indexed based on a simple cubic perovskite structure using space group *Pm3m*. The fundamental perovskite reflections shift to the left with increasing x, indicating increasing lattice parameters. Figure 1b illustrates the changes in the lattice parameters of (x)PIN-(1-x)PMN with increasing PIN. The relationship between the lattice parameters and the compositions is almost linear.

Pyrochlore peaks, identified with "*" in Fig. 1a, were found in the samples with x = 0.8 and x = 0.9. The presence of the pyrochlore phases in these two sintered samples reduced their densities. The intensity of the pyrochlore peaks in x = 0.8 is lower than that in x = 0.9, indicating that the presence of PMN in the solid solution improves the structural stability of PIN perovskite phase by its tolerance factor and electro-negativity [18]. The superstructure reflection 1/2 1/2 1/2 (*Pm3m*) cannot be observed in Fig. 1a, despite the fact that some of the ceramics contain higher amounts of PIN. The degree of the structural long-range 1:1 order could not be detected by x-ray diffractometry, indicating a very low degree of long-range order across the entire compositions.

**3.2 Relative Permittivity**

The $\varepsilon_r$ and tan $\delta$ of the (x)PIN-(1-x)PMN ceramics with x = 0.1 to 0.9 were measured. The range of the measuring frequencies was from 100 Hz to 2 MHz and the temperatures were from -30 ºC to 200 ºC. The temperature and frequency dependence of $\varepsilon_r$ and tan $\delta$ for the selected ceramics with x = 0.1, 0.3, 0.4 and 0.6 are shown in Figs. 2a-d, respectively. The maximum values of $\varepsilon_r$ ($\varepsilon_{max}$) at 100 Hz of x = 0.1, 0.3, 0.4 and 0.6 are 6230, 4484, 3913 and 3635, respectively, and the maximum values of tan $\delta$ at 100 kHz for these compositions are less than 0.1. The higher values of $\varepsilon_r$ are located at the lower frequencies. The values of $\varepsilon_{max}$ decreased with increasing amount of PIN and the temperatures at $\varepsilon_{max}$ ($T_{max}$) also increased. Although the maximum values of tan $\delta$ exhibit temperature dependence as $\varepsilon_{max}$, the values remain similar at all the compositions. Strong frequency dispersions close to $T_{max}$ were observed. All the compositions except x = 0.9 show the characteristics diffuse phase transition. For the ceramic with x = 0.9, the departure from the relaxor behavior might be related to the presence of the pyrochlore phase in the ceramic. As shown in Fig. 2, additional dielectric anomalies for x = 0.3, 0.4 and 0.6 were observed above $T_{max}$. The anomalies occurred at 90 °C and frequency dispersions on both $\varepsilon_r$ and tan $\delta$ were also noted. However, the amplitudes in $\varepsilon_r$ were decreased as frequency increased.

Figure 3 shows the $\varepsilon_r$ and tan $\delta$ of the (x)PIN:(1-x)PMN ceramics with x = 0.1 to 0.9 measured at 1 kHz as a function of temperature. The significant differences between $\varepsilon_r$ of different compositions can been found below 100 °C. Two groups of behaviors can be identified. In the x = 0.1 and 0.2, the values of $\varepsilon_r$ are similar and the transitions are relatively sharp compared to the second group including the sample with x ≥ 0.3. A dramatic decrease of $\varepsilon_r$ from x = 0.2 to 0.3 illustrates a boundary for distinguishing between the two different groups. The ceramics with x ≥ 0.3 have broader phase transition and the values of $\varepsilon_r$ between $T_{max}$ and 100 °C show less sensitivity to temperature. Figure 3 clearly shows the presence of dielectric anomalies at higher temperatures in most of the compositions, especially apparent in the plots of tan $\delta$. The temperature of this transition did not change significantly with composition. This transition at higher temperature reveals stabilities in both the temperature and composition. The vertical offset of tan $\delta$ in x = 0.9 is caused by the electrical contributions from the pyrochlore or other second phases.

Figure 4 shows the changes in $\varepsilon_{max}$ and $T_{max}$ as a function of the composition. Increasing the quantity of the PIN caused a decrease in $\varepsilon_{max}$. The change in $\varepsilon_{max}$ from 0.1 to 0.2 is relatively small but a sudden drop was noted when x increased from 0.2 to 0.3. A plateau was also observed between the PIN quantities from 0.4 to 0.7. The differences between the values of $\varepsilon_{max}$ on the plateau were within ~10%. The significant difference between x = 0.8 and 0.9 was caused by the existence of the pyrochlore phase. Similar to the classification, the differences between the values of $\varepsilon_{max}$ and $\varepsilon_r$ at 25 °C were very close in the samples with x ≥ 0.3. However, the change of $T_{max}$ did not follow the tendency of $\varepsilon_{max}$. The temperatures of $\varepsilon_{max}$ were increased almost linearly as the PIN quantity increased.

In relaxors, the value of $\varepsilon_r$ above $T_{max}$ does not follow the Curie-Weiss law. A number of materials show the intermediate behavior between proper ferroelectric and relaxor. A variable power-law was developed to describe the intermediate behavior [19]. This power-law differs from the quadratic Curie-Weiss law by introducing an exponent γ as an empirical parameter, described as:

$$\frac{1}{\varepsilon_r} = \frac{1}{\varepsilon_{max}} + \frac{(T-T_{max})^\gamma}{2\varepsilon_{max}\delta^\gamma}$$

The value of γ can vary between one and two. For a perfect proper ferroelectric and a perfect relaxor the value of γ= 1 and γ = 2, respectively. The denominator is usually presented as a Curie-Weiss type constant, which is named as the modified Curie-Weiss constant (C'), is given by:

$$C' = 2\,\varepsilon_{max}\,\delta^\gamma$$

Although the above equations were modified, δ is still an effective empirical value to describe the width of the diffuse phase transition [20]. These values of δ and γ can be obtained by a log-log plot of $(1/\varepsilon_r - 1/\varepsilon_{max})$ against $(T-T_{max})$. The slope of the curve represents γ and the value of δ is determined by the intercept Y at the y-axis, as:

$$\delta = [\frac{\exp(-Y)}{2\varepsilon_{max}}]^{1/\gamma}$$

In order to identify the quality of the fit, the linear regression coefficient, $R^2$, is shown see Table 1. Other important parameters of the dielectric properties of these ceramics with x = 0.1 to 0.9 are also listed in Table 1. The empirical exponents of γ varies from 1.22 to 1.59 in different compositions. A particular compositional dependence could not be determined. Qualitatively, the γ-values of the samples with lower PIN quantity is slightly higher than the values found in the sample with higher x. These ceramics are neither perfect ferroelectric nor perfect relaxor but the observations are highly correlated with the order-disorder nature of the end members. The diffusiveness (δ) and the modified Curie constants (C') also vary with the compositions x= 0.1 to 0.9. As in the previous classification, the values of δ in the ceramics with x ≥ 0.3 are varied from 100 to 175 °C. In contrast, those values in the ceramics with x < 0.3 did not change significantly and are about 75 °C only. The largest δ found in the ceramic with x = 0.9 could be caused by the presence of the pyrochlore phase.

In addition to using modified Curie-Weiss law, the relaxation features in a relaxor can be characterized by the Vogel-Fulcher (V-F) relationship [21]: $\nu = \nu_o \exp[-E_a/k_B(T_{max}-T_o)]$, where $\nu$ is the measured frequency, $k_B$ is the Boltzmann's constant, $\nu_o$ is the attempt (also termed Debye) frequency, $E_a$ is the activation energy, and $T_o$ is the freezing temperature. $\nu_o$, $E_a$ and $T_o$ are phenomenological parameters, which are obtained by curve fitting. Figure 5 shows a plot of $\ln(\nu)$ as a function of $T_{max}$ for a selected group of (x)PIN:(1-x)PMN ceramics with x= 0.1, 0.3, 0.4 and 0.6. The experimental data are illustrated with symbols and the curve fittings to the data are shown with solid lines. In Figure 5, the non-linear relationship between the frequency and temperature of the $\varepsilon_{max}$ indicates that the experimental data do not follow the simple Debye equation.  However, this non-linear relationship can be fitted quite satisfactorily using the V-F relationship, which verifies the characteristic of a relaxor. The phenomenological parameters, $\nu_o$, $E_a$ and $T_o$ for (x)PIN:(1-x)PMN were determined by fitting the V-F relationship and are listed in Table 2. The measurements show fluctuations in the values of $\nu_o$, $E_a$ and $T_o$ without any particular order in the fluctuations with changing composition, as shown in table 2.  This observation is analogous to the case of the exponent γ.  In the compositions with x = 0.1 to 0.8, the values of $E_a$ fluctuates from 0.0380 to 0.06 eV, which are smaller compared to PMN [21]. The values of $T_o$ show inclination to increase with increasing quantity of PIN, which is also analogous to the case of $T_{max}$. In contrast, dependence of $\nu_o$ on compositional is more complex, primarily because the values lying between 0.1 to 0.4 are within $3.5 - 5.0 \times 10^{10}$Hz, whereas the values between 0.5 to 0.8 are $1.5 \times 10^{10}$Hz. These significant fluctuations are not easy to elucidate.  The ceramics with x = 0.9, is not included in these studies because it has been found to contain a significant amount of pyrochlore phase.

**3.2 Hysteresis loop**

Ferroelectric hysteresis loops of the (x)PIN:(1-x)PMN ceramics were measured at room temperature at a frequency of 10 Hz. Figures 6a-d show the hysteresis loops of the ceramics with x

= (a) 0.1, (b) 0.3, (c) 0.6 and (d) 0.8, respectively. At the applied electric field of 4 kV/mm, the maximum values of the remanent polarization ($P_r$) and the coercive field ($E_c$) are only < 2.0 $\mu C/cm^2$ and < 0.3 kV/mm, respectively. Negligible hysteresis loops were obtained in these compositions, indicating the absence of macroscopic ferroelectric domains throughout the ceramics.

Figure 7 shows the hysteresis loops of the ceramics with x = 0.1 to 0.9 measured at 4 kV/mm. All compositions reveal slim hysteresis loops, excluding x = 0.9. The hysteresis loops and the corresponding $P_r$ and $E_c$ values of the ceramics with x = 0.1 to 0.7 did not vary significantly. The linear decrease in the maximum electric displacement ($D_{max}$) originates from an increase in the PIN quantity. The decreases in $D_{max}$ were relatively large in ceramics with x = 0.8 and 0.9. In the x=0.9 ceramic, the more rounded and open hysteresis loop is the consequence of the electric conduction through the sample, related to the higher dielectric loss as shown in tan δ (figure 3).

The values of $P_r$, $E_c$ and $D_{max}$ of the ceramics measured at 4 kV/mm are listed in Table 3. Based on the results of the $P_r$ and $E_c$, it can be quantitatively concluded that all the ceramics show negligible amount of hysteresis. The values of $D_{max}$ depend on compositions, although the differences in compositions are small. The higher $D_{max}$ can be obtained in the ceramics with lower PIN quantity. Combined with the previously shown dielectric properties, (x)PIN:(1-x)PMN ceramics, therefore, are relaxors instead of normal ferroelectrics with diffuse phase transitions.

**3.3 Structural long-range ordered domains**

Centered dark-field (CDF) images of the 1:1 structural long-range ordered (LRO) domains in the as-sintered ceramics with x=0.1, 0.3, 0.4 and 0.6 are shown in Figs. 8a-d, respectively. The images were taken using one of the superstructure reflections {1/2 1/2 ½} that yield domains of high contrast corresponding to the regions having the 1:1 LRO structural ordering. In the ceramic (0.1)PIN:(0.9)PMN, the size of the ordered domains varies from 3 nm to 30 nm across, which are slightly larger than those observed in PMN. When the PIN quantity is increased, the size of 1:1

LRO domains also increased. Similar to many order-disorder complex perovskite ceramics, such as Pb(Sc$_{1/2}$Ta$_{1/2}$)O$_3$ [22], the degree of ordering is inhomogeneous near grain boundaries where the domains are larger than those located at the centre of the grains or away from the boundaries. This kind of spatially preferential ordering can be observed in the ceramics with x ≥ 0.3, figures 8b-d. These images show that the ceramics have intermediate degrees of 1:1 B-site cation ordering. In the (0.6)PIN:(0.4)PMN ceramic, some large ordered regions, ~200 nm across, are found (figure. 8d). Extended anti-phase boundaries are clearly seen reminiscent of anti-phase boundaries observed in some highly ordered complex perovskites [23]. However, the CDF image shows that these ordered regions are not composed of single domains (figure 8e). In addition to the distinct anti-phase boundaries, numerous fine boundaries, which separate nano-size structural ordered domains, are also observed.

## 4. Discussion

The dielectric properties of (x)PIN:(1-x)PMN show compositional dependence. Both values of ε$_r$ and D$_{max}$ decrease with increasing PIN quantity in the solid solution. However, there is no anomalous enhancement of the properties in a particular composition. As a result, morphotropic phase boundary does not exist in this relaxor-relaxor solid solution. The ceramics with x = 0.3 to 0.7 show their stability of ε$_r$ in a range of temperatures. These as-sintered ceramics have already fulfilled the EIA-standard [24] as Z6P ceramic dielectrics in which the variation of ε$_r$ from 10 °C to 105 °C is ±10 % compared to ε$_r$ at 25 °C. Some compositions of (x)PIN:(1-x)PMN with their high ε$_r$ and broad phase transition of the ceramics could be good candidates for capacitor applications. The existence of the higher temperature transition in (x)PIN:(1-x)PMN is similar to some disordered complex perovskites [23] or even irradiated copolymers [25,26]. This transition is highly related to the transformation between spontaneous relaxor ferroelectric to normal ferroelectric, which originates from the hetero-phases fluctuations or micro- to macro-domains transition [27]. It is believed that hetero-phase fluctuation possibly occurs in (x)PIN:(1-x)PMN because of the evidence of nano-size ordered domains and disordered regions observed in the TEM. In this study, we show that a disordered ceramic undergoes high-temperature transition, which is diminished by partial annealing; and entirely removed by fully annealing the ceramic [28]. The indications are that the transition behaviour in (x)PIN:(1-x)PMN is highly influenced by the degree of 1:1 structural long-range order.

In (x)PIN:(1-x)PMN, the correlation between the electrical properties and 1:1 structural ordered domains cannot be explained simply by the conventional order-disorder description in complex perovskites. The observed relaxor behavior of (x)PIN:(1-x)PMN ceramics in which ordered domains have similar morphology to an intermediate ordered material, such as x = 0.6, cannot directly correlate to the usual order-disorder model. In fact, a number of complex perovskite solid solutions also deviate from such order-disorder behavior. For example, in $Pb(Mg_{1/3}Ta_{2/3})O_3:PbZrO_3$ (PMT-PZ) solid solution [29], the frequency dispersion in relative permittivity is still present in the annealed ceramics in which the size of ordered domains was significantly enhanced.

In the TEM study of (x)PIN:(1-x)PMN, the size of the ordered domains increase with increasing PIN quantity. The enhancement of the degree of 1:1 structural LRO continues from x = 0.1 to 0.9, although x-ray diffractometry cannot reveal the superlattice reflection readily. In other studies, a similar order-disorder PMN-based solid solution, $(x)Pb(Sc_{1/2}Nb_{1/2})O_3:(1-x)Pb(Mg_{1/3}Nb_{2/3})O_3$, abbreviated to (x)PSN:(1-x)PMN, shows a saturation of the long-range order within a particular compositional regions [30]. The as-sintered (0.5)PSN:(0.5)PMN has both the highest degree of 1:1 long-range ordering and the largest ordered domain size. But there is no such limit in (x)PIN:(1-x)PMN to terminate the domain growth although both (x)PSN:(1-x)PMN and (x)PIN:(1-x)PMN also belong to $Pb(B^{3+}Nb^{5+})O_3:PMN$ solid solutions.

The large ordered domains, in fact, were themselves composed of nano-size ordered domains. This observation has not been reported in (x)PSN:(1-x)PMN and other Pb-based perovskite solid solutions. No superlattice reflection on the x-ray diffraction patterns are observed in all the compositions. We propose that the coherence of the long-range structural ordering in (x)PIN:(1-x)PMN was broken. The B-site cation ordering was locally formed in either nano or sub-nano scale, which has been shown in the dark-field images. Because of the shorter coherence length of electrons, the dark-field technique in TEM enables ordered regions to be imaged, similar to the case of PMN. The three different B-site cations create different ordered regions separated by fine boundaries may well be caused by mixed-ordering [13]. The experimental results and the simulation of mixed ordering in different Pb-based complex perovskites were published elsewhere [31].

By combining the results of the influence on dielectric properties and the existence of nano-scale ordered domains, it is believed that random field model [32,33] is a suitable approach to describing the relaxor behavior in (x)PIN:(1-x)PMN. The consideration of long-range structural ordering cannot alone directly explain the relaxor properties observed in the (x)PIN:(1-x)PMN ceramics. However, the different size and quantity of the nano-size domains and the short-range order can be considered as the fluctuations of hetero-phases, consequently a possible source of the random fields.

## 5. Conclusions

(x)PIN:(1-x)PMN solid solution were fabricated and characterized. Pure perovskite phase was obtained in the ceramics with x ≤ 0.7 by the conventional mixed oxide route. It is confirmed that all the as-sintered ceramics exhibit relaxor behavior. The relative permittivity and electric displacements are diverse and depend on compositions. The verification of the dielectric results by a Vogel:Fulcher type relationship proves that the ceramics of (x)PIN:(1-x)PMN are relaxors.

The decreases of $T_{max}$ and $D_{max}$ are almost linear when the PIN quantity is increased. Morphotropic phase boundary does not exist in this solid solution. However, the changes in $\varepsilon_{max}$ are also diverse but a plateau was found in compositions between x = 0.4 to 0.7. In the TEM study, the size of the structural ordered domains is enhanced by increasing PIN quantity. But, in the ceramics containing high PIN quantity, it is found that the large ordered domains separated by anti-phase boundaries are themselves composed of numerous individual nano-sized ordered domains. Although the degree of 1:1 order increases with increasing values of x, the relaxor ferroelectric behavior is still maintained. The consideration of the properties discussed in this paper, make the ceramics with x = 0.3 to 0.7 good candidates for capacitor applications with satisfactory temperature stability.

**Acknowledgements**

C.W. Tai is grateful for the financial support by grant B-Q891.

Table 1. Dielectric properties of (x)PIN:(1-x)PMN ceramics with x = 0.1 to 0.9.

| Composition (x) | $T_{max}$ (°C) | $\varepsilon_{max}$ | $\varepsilon_r$ at 25 °C | $\gamma$ | $\delta$ (°C) | C' (x10$^5$ °C) | $R^2$ |
|---|---|---|---|---|---|---|---|
| 0.1 | -1.7 | 6230 | 5658 | 1.45 | 73.2 | 61.86 | 0.9972 |
| 0.2 | 2.9 | 5946 | 5492 | 1.59 | 73.9 | 119.77 | 0.9984 |
| 0.3 | 7.3 | 4484 | 4373 | 1.51 | 142.9 | 165.19 | 0.9878 |
| 0.4 | 11.2 | 3913 | 3833 | 1.47 | 100.4 | 70.60 | 0.9850 |
| 0.5 | 16.1 | 3621 | 3579 | 1.49 | 118.2 | 90.68 | 0.9895 |
| 0.6 | 20.6 | 3635 | 3629 | 1.28 | 138.3 | 42.55 | 0.9873 |
| 0.7 | 27.1 | 3449 | 3436 | 1.22 | 172.5 | 38.36 | 0.9883 |
| 0.8 | 31.6 | 3027 | 2928 | 1.32 | 105.3 | 29.57 | 0.9992 |
| 0.9 | 37.4 | 1346 | 1272 | 1.22 | 242.3 | 24.22 | 0.9971 |

Table 2. Summary of the phenomenological parameters in the Vogel-Fulcher relationship of the (x)PIN:(1-x)PMN ceramics with x = 0.1 to 0.8.

| Composition (x) | $\nu_o$ (x 10$^{10}$Hz) | $E_a$ (eV) | $T_o$ (K) | $R^2$ |
|---|---|---|---|---|
| 0.1 | 5.39 | 0.0469 | 241.2 | 0.9945 |
| 0.2 | 3.87 | 0.0499 | 242.6 | 0.9948 |
| 0.3 | 4.15 | 0.0602 | 240.1 | 0.9964 |
| 0.4 | 4.34 | 0.0476 | 248.4 | 0.9900 |
| 0.5 | 1.44 | 0.0444 | 257.4 | 0.9976 |
| 0.6 | 1.33 | 0.0435 | 261.1 | 0.9910 |
| 0.7 | 1.50 | 0.0380 | 272.5 | 0.9952 |
| 0.8 | 1.70 | 0.0572 | 266.7 | 0.9948 |

Table 3. Summary of ferroelectric hysteresis loop measurements of the (x)PIN:(1-x)PMN ceramics with x = 0.1 to 0.9.

| Composition (x) | $E_c$ (kV/mm) | $P_r$ (µC/cm$^2$) | $D_{max}$ (µC/cm$^2$) |
|---|---|---|---|
| 0.1 | 0.19 | 1.83 | 19.3 |
| 0.2 | 0.21 | 1.50 | 19.1 |
| 0.3 | 0.28 | 1.51 | 18.7 |
| 0.4 | 0.28 | 1.71 | 18.3 |
| 0.5 | 0.31 | 1.52 | 18.1 |
| 0.6 | 0.21 | 0.93 | 17.7 |
| 0.7 | 0.16 | 0.79 | 16.1 |
| 0.8 | 0.19 | 0.80 | 11.9 |
| 0.9 | 0.89 | 1.86 | 7.66 |

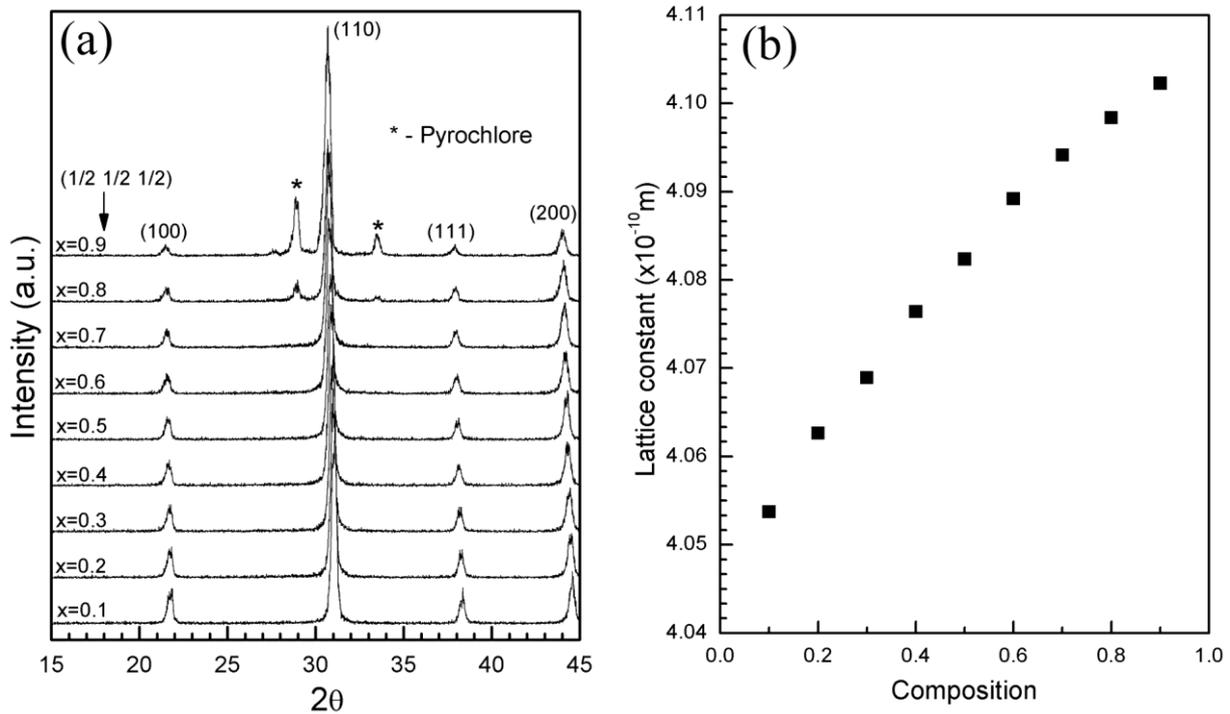

Fig. 1. (a) X-ray diffraction patterns of the (x)PIN:(1-x)PMN with x = 0.1 to 0.9. Simple perovskite phase is indexed as cubic. Pyrochlore phase are marked as "*". The corresponding peak position of the superlattice reflection 1/2 1/2 1/2 is also indicated by an arrow but no 1/2 1/2 1/2 peak has been observed. (b) Lattice parameters against different compositions. The lattice parameter almost linearly increases as increasing x.

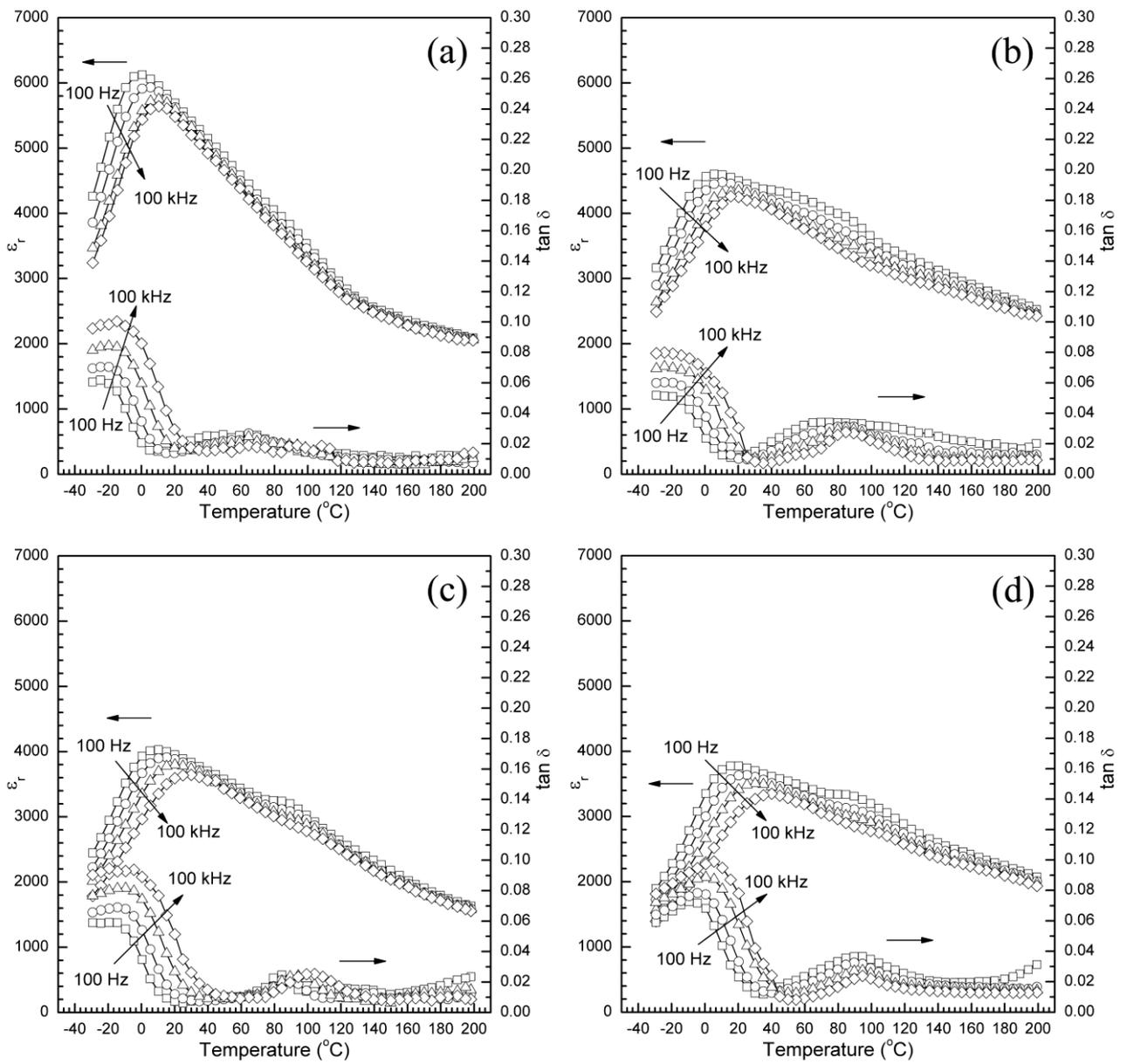

Fig. 2. Temperature and frequency dependence of $\varepsilon_r$ and tan $\delta$ of (x)PIN:(1-x)PMN ceramics with x = (a) 0.1, (b) 0.3, (c) 0.4 and (d) 0.6.

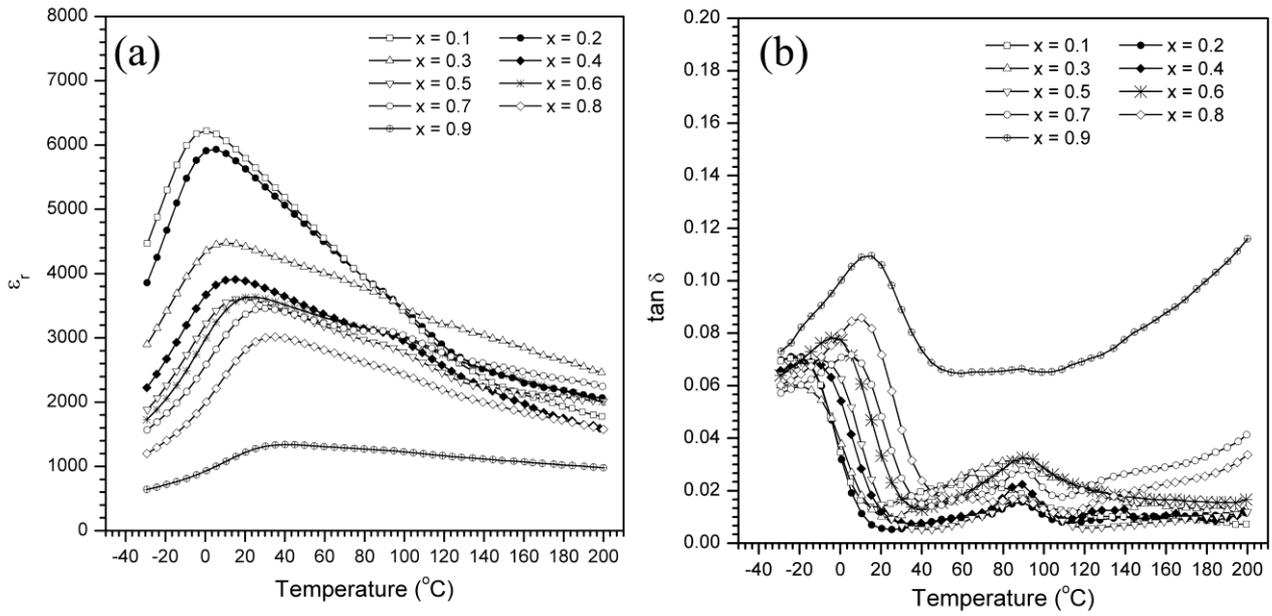

Fig. 3. Temperature dependence of $\varepsilon_r$ and tan $\delta$ for (x)PIN:(1-x)PMN ceramics with x = 0.1 to 0.9.

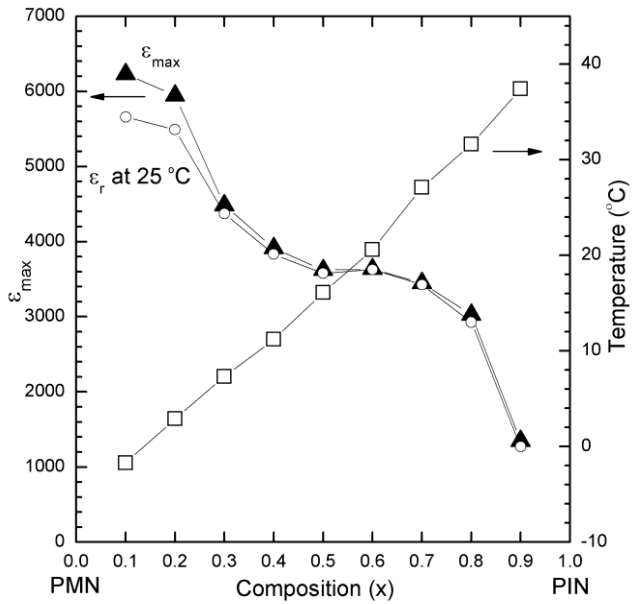

Fig. 4. Variation in $\varepsilon_{max}$ and the corresponding $T_{max}$ of (x)PIN:(1-x)PMN as a function of the composition.

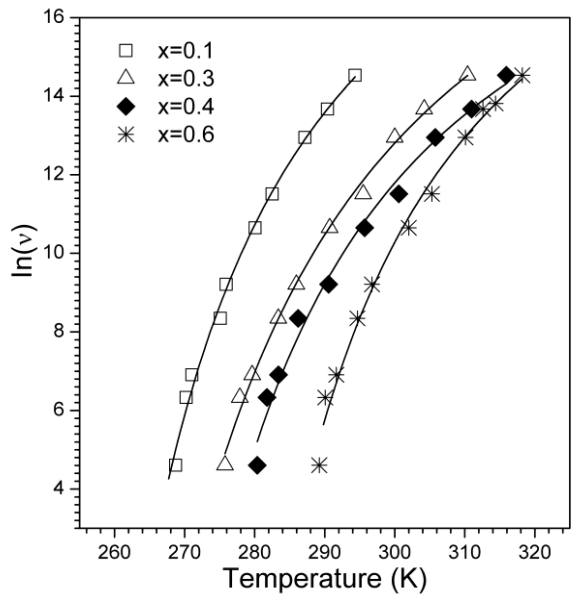

Fig. 5. Plot of In(ν) as a function of $T_{max}$ for (x)PIN:(1-x)PMN ceramics with x = (a) 0.1, (b) 0.3, (c) 0.4 and (d) 0.6. The symbols are the experimental data and the solid lines are the fitted curve.

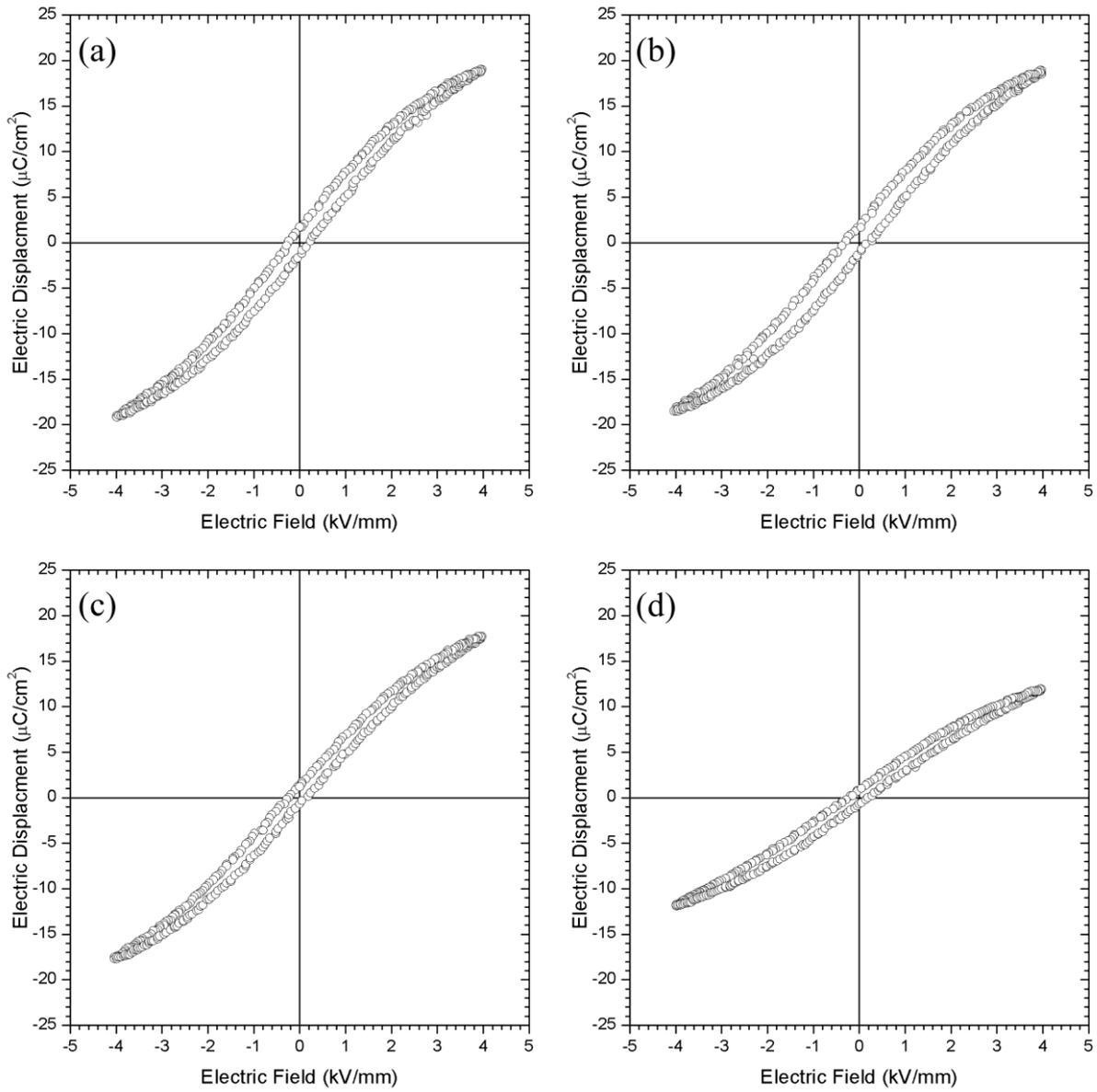

Fig. 6. Ferroelectric hysteresis loops measured at different electric fields for (x)PIN:(1-x)PMN ceramics with x = (a) 0.1, (b) 0.3, (c) 0.6 and (d) 0.8.

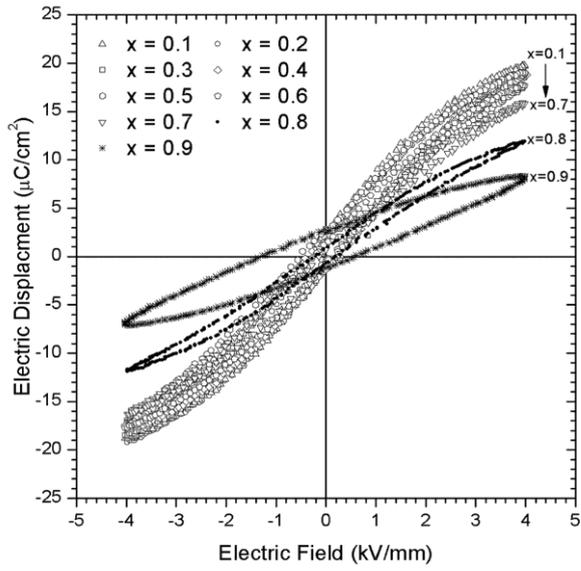

Fig. 7. Ferroelectric hysteresis loops of (x)PIN:(1-x)PMN ceramics with x = 0.1 to 0.9 measured at an applied electric field of 4 kV/cm.

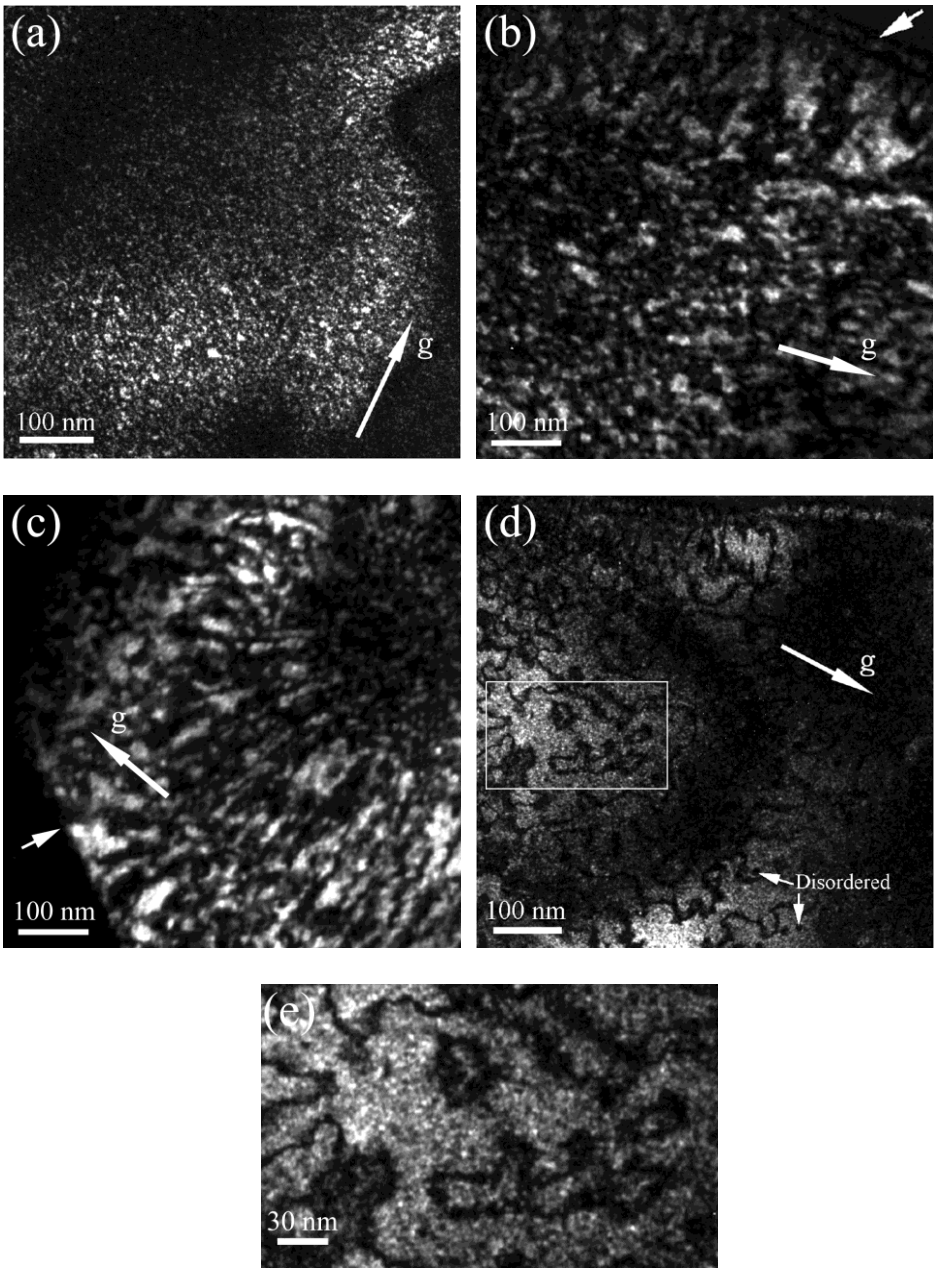

Fig. 8. Dark field micrographs g=1/2 1/2 1/2 of as-sintered (a) (0.1)PIN:(0.3)PMN, (b) (0.3)PIN:(0.7)PMN, (d) (0.4)PIN:(0.6)PMN and (d) (0.6)PIN:(0.4)PMN ceramics. In Fig. b and c, preferential ordering is observed near the grain boundaries indicated by arrows. In Fig. d, the ordered domains and the anti-phase boundaries are clearly visible. Some disordered regions are indicated by arrows. (e) Enlarged CDF image of the selected area in Fig. d. Nano-size regions, which have different contrast, are observed in the ordered regions.